\documentclass[pra,aps,twocolumn,nofootinbib,floatfix]{revtex4}
\usepackage{graphicx}

\newcommand{\dpar}[2]{\frac{\partial#1}{\partial#2}}
\def\be{\begin{equation}}
\def\ee{\end{equation}}
\def\ba{\begin{eqnarray}}
\def\ea{\end{eqnarray}}

\def\part{\partial}

\def\gam{\gamma}
\def\D{\Delta}

\def\nref{n_{\rm ref}}
\def\tV{\widetilde{V}}
\def\tJ{\widetilde{J}}
\def\cR{{\cal R}}
\def\cI{{\cal I}}
\def\cN{N}
\def\mic{\mbox{$\mu$m}}
\begin{document}
\title{Acoustic horizon as a phase-slip surface}
\author{Sergei Khlebnikov} 
\affiliation{Department of Physics and Astronomy, Purdue University, West Lafayette, IN 47907, USA}


\begin{abstract}
A recent experiment has demonstrated formation of a supersonic region in a convergent 
two-dimensional flow of a condensate of cesium atoms.
Theoretical description of this effect has made use of
stationary solutions to the Gross-Pitaevskii equation
with a 3-body dissipative term.
Here, we further develop that description, focusing on a
new stationary solution, linear stability analysis, and properties
of the time-dependent ``resistive" state. 

\end{abstract}

\maketitle

\section{Introduction}

Loss of superfluidity at a critical current is a classic problem, theoretical study of which 
goes back to Landau's work on a stability criterion \cite{Landau}. 
That work has emphasized dependence of stability of the flow on 
the relation between the speed of the flow and
phase velocities of excitations that can propagate in the fluid, such as the sound.
In a separate development, supersonic flows have been proposed \cite{Unruh} as a means 
of creating an acoustic horizon, analogous to the event horizon of a black hole, with the 
purpose of simulating Hawking radiation predicted \cite{Hawking} for the latter. By now,
a significant experimental effort has been directed towards 
realizing this physics in one-dimensional (1D) superflows \cite{Steinhauer,de_Nova&al,Kolobov&al}, 
and it seems there is a consensus that  the
instability associated with the Landau criterion does play a role in this case, at stages following
formation of an inner acoustic horizon \cite{Kolobov&al,Tettamanti&al,Wang&al}.

Recently, an experiment by the Purdue group \cite{self-osc} has obtained a convergent 
two-dimensional (2D) supersonic flow in a superfluid of cesium atoms. 
This work has observed 
self-induced oscillations in the flow, which the analysis has
attributed to complex nonlinear dynamics associated with emission of ring
solitons from the supersonic region (even though individual solitons could not be 
identified in the experiment due to limits of the spatial resolution). Parallels have been drawn 
with two phenomena discussed in the earlier literature: continuous emission of solitons by 
a superfluid flowing over an obstacle in one dimension 
\cite{Hakim,Pavloff,Engels&Atherton}, on the one hand, and 
formation of a localized self-oscillating state---a phase-slip center \cite{SBT}---in a 
superconducting wire, on the other. (For two-horizon states, emission of solitons trains in one
dimension has been
studied numerically in Refs.~\cite{Michel&Parentani,deNova&al:2016}.)

The two phenomena just mentioned are in a sense at the opposite ends of the inertia 
versus dissipation spectrum. Indeed, the local dynamics described by the conservative
1D Gross-Pitaevskii (GP) equation, on which calculations of soliton emission \cite{Hakim,Pavloff,Michel&Parentani,deNova&al:2016} have been based,
is purely inertial, the only dissipative effect being
radiation of solitons and sound to infinity. In contrast, dynamics at the phase-slip center
of Ref.~\cite{SBT} involves local dissipation, due to a diffusive normal component. 
The convergent 
radial superflow engineered in Ref.~\cite{self-osc} has both types of dissipative mechanisms, 
with local loss of
particles provided by three-body recombination at the center of the trap. For this reason, 
the system is of special interest as a platform for studying the combined effect of dissipation 
and nonlinearity.

Here, after a brief review (in Sec.~\ref{sec:gs}) of the theoretical methods used 
in our previous work \cite{self-osc}, 
we present two additional results concerning the bifurcation that has been a part 
of the description proposed there. First, we observe that, even after the superfluid 
ground state
disappears through a saddle-node bifurcation, there remains a second superfluid solution, which 
can be viewed as a bound state of a ring soliton 
in the central region. The domain of attraction of this second solution, though, appears to be
quite small, and a quench to a current not far above the critical leads already 
to the complex quasi-periodic dynamics described in Ref.~\cite{self-osc} 

Second, we propose a numerical diagnostic
of that complex dynamics based on the rate of change of the phase of the order parameter in the 
central region. While in a superfluid state the phase is a constant or performs periodic
oscillations, in the quasi-periodic regime, it grows without limit at a more or less constant rate 
(with a quasi-periodic component superimposed). This strengthens the analogy between the present system
and the resistive state of superconducting wires (where, by Josephson's relation, the rate of 
change of the phase is proportional to the voltage). A further discussion of the results
is given in the concluding section.

\section{The ground state and the droplet}
\label{sec:gs}

As in Ref.~\cite{self-osc}, our starting point is the 2D classical GP equation with 
an additional 3-body dissipative term
\be
i \hbar \part_t \psi = \frac{\hbar^2}{2 m} \left( - \nabla^2 + 2 g n - i \gam_3 n^2 \right) \psi
+ [V(r) - \mu ] \psi \, ,
\label{eq}
\ee
where $n = \psi^\dagger \psi$ in the 2D atom density, $r$ is the radial coordinate, and
$g$ and $\gamma_3$ are real positive constants describing the pairwise atomic interaction and 3-body
dissipation, respectively. The dissipative coefficient $\gamma_3$ is related to the 3-body 
recombination rate $L_3$ in $\dot{n} = - L_3 n^3$ by $\gam_3 = m L_3 / \hbar$ 
(with $2m / \hbar = 4.185$ ms/$\mic^2$ for cesium). The addressing potential
$V(r)$ is taken in the Gaussian form, as
\be
V(r) = - V_0 \exp( -2 r^2 / r_s^2 ) 
\label{pot}
\ee
with $V_0 > 0$. This potential causes the particles to flow towards the center of the trap, where as
a consequence the density becomes high, and the particles are 
efficiently removed by 3-body recombination.

As far as solutions to the nonlinear problem (\ref{eq}) go, we consider only those with perfect 
rotational symmetry, i.e., $\psi(r, t)$. On the other hand, when studying linear stability of
a stationary solution so obtained, 
we allow for arbitrary eigenmodes of the form $\chi(r) e^{i\ell \phi}$,
where $\phi$ is the azimuthal angle, and $\ell$ an integer. This amounts to the replacement
\be
\nabla^2 = \frac{1}{r} \dpar{}{r} \left( r \part_r \right) - \frac{\ell^2}{r^2} 
\label{repl}
\ee
in the linearized version of the problem.

Eq.~(\ref{eq}) is supplemented by the standard regularity condition $\part_r \psi = 0$ 
at $r=0$ and a boundary condition at an outer radius $r = R$. We wish to
choose the condition at $r = R$ so as
to allow, at least in principle, for existence of stationary solutions, in which the loss of
particles at the center is precisely compensated by inflow from the boundary. This is an
idealized version of the situation when there is a flow of particles
from large radii, resulting in a slowly-changing (albeit not precisely stationary) state.
A convenient choice is \cite{self-osc}
\be
\left. \part_r \cR(r,t) \right|_{r=R} = 0 \, , \hspace{3em} \left. \cI(r,t) \right|_{r=R} 
= 0 \, ,
\label{bc}
\ee
where $\cR$ and $\cI$ are, respectively, the real and imaginary parts of $\psi$. Note the
different boundary conditions for $\cR$ and $\cI$. The conditions (\ref{bc}) 
set the phase of $\psi$ at $r = R$ to zero and, as can be seen by computation
of the particle current, allow for a nonzero particle inflow at the boundary.
Alternatively, one can observe that these conditions break the symmetry $\psi \to e^{i\alpha} 
\psi$, where $\alpha$ is a real parameter; this is the symmetry responsible,
via Noether's theorem, for conservation of the particle number. 
Eq.~(\ref{bc}) however does not determine the amount of the inflow, which is then found as
a part of the solution \cite{self-osc}.

The depth $V_0$ of the potential controls the speed of the flow and is considered a variable.
The other parameters are fixed at values that are close to those in Ref.~\cite{self-osc} 
and are listed in 
Table~\ref{param}. Note that $\gam_3$ is specified in terms of $L_3$ as described above, and the 
chemical potential $\mu$ in terms of a reference density $\nref$ as $\mu = \hbar^2 g \nref / m$.

\begin{table}
\begin{center}
\begin{tabular}{|c|c|c|c|c|}
\hline
$g$ &  $r_s$ ($\mic$) & $R$ ($\mic$) & $\nref$ ($\mic^{-2}$) & $L_3$ ($\mic^4$/ms) \\
\hline
0.42 & 6.5 & 26 & 9.9 & $4.3\times 10^{-5}$ \\
\hline
\end{tabular}
\end{center}
\caption{Values of the parameters used for numerical calculations.}
\label{param}
\end{table}

We discretize the right-hand side of (\ref{eq}) on a uniform grid of $N$ points and separate its
real and imaginary parts $\hbar F_R$ and $\hbar F_I$, 
to write (\ref{eq}) as a system of $2N$ algebraic equations
\ba
\part_t \cR_j & = & F_{Ij} (u_i) \, , \label{eqR} \\
\part_t \cI_j & = & -F_{Rj} (u_i) \, , \label{eqI} 
\ea
where the $2N$-dimensional vector $u_i$ is composed of $\cR_j$ and $\cI_j$, as follows
\be
u_j = \cR_j \, , \hspace{3em} u_{j + \cN} = \cI_j \, ,
\label{uj}
\ee
for $j = 0, \dots \cN - 1$. The components $F_{Rj}$ and $F_{Ij}$, on the other hand, can be combined
either into a vector $V_i$, as
\be
V_j = F_{Rj} \, , \hspace{3em} V_{j + \cN} = F_{Ij} \, ,
\label{Vj}
\ee
or into a different vector $\tV_i$, as
\be
\tV_j = F_{Ij} \, , \hspace{3em} \tV_{j + \cN} = - F_{Rj} \, .
\label{tVj}
\ee

Two different vector fields defined above give rise to two different Jacobians.
One is 
\be
J_{ik} = \dpar{V_i}{u_k} 
\label{J}
\ee
and is useful for studying stationary solutions---those for which $V_i = \tV_i = 0$. 
Indeed, consider a stationary solution to the dissipative problem (\ref{eq}) but look first at
the Jacobian $J$ computed in the absence of dissipation, i.e., for $\gam_3 = 0$. That Jacobian
is a symmetric matrix and so has only real eigenvalues. If an eigenvalue is simple, we may then
expect it to remain real when the dissipation is nonzero but not too strong. 
This is convenient if one wishes to study merging of two stationary solutions at a 
bifurcation point. The other Jacobian,
\be
\tJ_{ik} = \dpar{\tV_i}{u_k} \, ,
\label{tJ}
\ee
is necessary for linear analysis of stability of solutions. Its eigenvalues are either real or
come in complex conjugate pairs. 

\begin{figure}
\begin{center}
\includegraphics[width=3.25in]{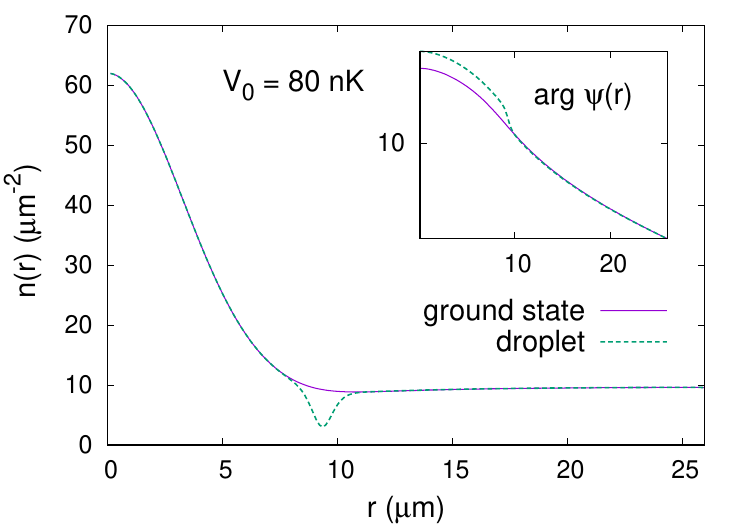}
\end{center}
\caption{\small
Profiles of the density and the phase of $\psi$ (inset) for two stationary solutions 
that exist when the potential depth $V_0$ is below critical.}
\label{fig:two_solutions}
\end{figure}

Examples of stationary solutions, obtained by applying the multidimensional Newton-Raphson (NR) 
method to the discretized problem with parameters listed in Table~\ref{param}, 
are shown in Fig.~\ref{fig:two_solutions}. Linear algebra required
by that method, as well as computation of the eigenvalues of $J$ and $\tJ$, has been done using
the \textsc{lapack} library \cite{lapack}. For finding stable ground states, there 
is an alternative to the
NR method---computing the endpoint of real-time evolution with a reasonable initial condition; 
results from these different methods agree.

The two solutions shown in Fig.~\ref{fig:two_solutions}
correspond both to the same value of the potential depth $V_0$. For the solution
with a dip in the density profile, the Jacobian $J$ has a single
negative eigenvalue for each of several values of the orbital number $\ell$, including
$\ell = 0$. Using terminology already in place for resistive transitions in 
superconducting wires \cite{Little,LA}, we refer to this solution as the droplet. 
For the other, smoother solution, $J$ has no negative 
eigenvalues, and we refer to it as the ground state. 

As we increase $V_0$, the solutions merge and disappear via a bifurcation at a critical 
$V_0 = V_{c1}$ (here $V_{c1} \approx 87.8$ nK). 
The critical solution has
a small supersonic region, with an inner horizon at $r\approx 9~\mic$ and an outer horizon at
$r\approx 10~\mic$. For this reason, the loss of the superfluid ground state at the bifurcation 
point has been interpreted in Ref.~\cite{self-osc} as an expression of the Landau criterion.

The results in this section and the next---with the exception of Fig.~\ref{fig:soliton} 
(inset)---are from a grid of size $N=200$, which is smaller
than $N=500$ used in Ref.~\cite{self-osc} Where a comparison has been made, 
however, the results from $N=200$ and $N=500$ are quite similar. For example, increasing the grid
size to $N=500$ shifts the critical potential depth $V_{c1}$ by a fraction of a percent, 
to $V_{c1} \approx 88.0$ nK, while changes in the profiles of the ground state and droplet for
$V_0 = 80$ nK,
relative to those in Fig.~\ref{fig:two_solutions}, are almost imperceptible by eye. The inset of 
Fig.~\ref{fig:soliton} in the next section
presents a comparison of results from two grid sizes for another type of solution, whose
radial dependence is particularly sharp.

Linear stability analysis, based on eigenvalues of the Jacobian $\tJ$, shows that the droplet 
always  has an unstable mode with $\ell = 0$. Away from the critical point, it also has unstable 
modes with nonzero $\ell$ but those disappear as $V_0$ gets closer to $V_{c1}$. 
At the critical point, the (real) eigenvalue corresponding to the unstable mode approaches zero, 
confirming that the bifurcation is of the saddle-node type.

\section{The second superfluid solution}
\label{sec:sfsol}

\begin{figure}
\begin{center}
\includegraphics[width=2in]{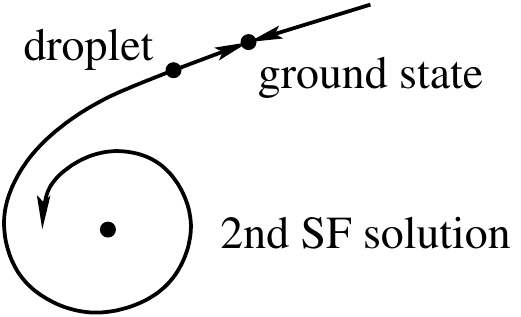}
\end{center}
\caption{\small
A schematic diagram illustrating decay of the droplet to the ground state or to another 
superfluid (SF) state, depending on the direction of the initial displacement. In this example, 
the second SF state
corresponds to a limit cycle near an unstable stationary solution.}
\label{fig:portr}
\end{figure}

It is instructive to follow the nonlinear rotationally-symmetric
evolution, as described by Eq.~(\ref{eq}), starting with the field
displaced from the droplet along the unstable mode. 
We find different dynamics depending on the direction 
of the displacement. For one of the directions, the field relaxes to the ground-state solution, but
for the other it approaches a different superfluid state, which is either a new stationary 
solution or a small limit cycle near such. For $V_0 = 80$ nK (the same value as used for
Fig.~\ref{fig:two_solutions}), the second option is realized, so the relevant orbits
in the rotationally-symmetric problem look schematically as shown in 
Fig.~\ref{fig:portr}. 

\begin{figure}
\begin{center}
\includegraphics[width=3.25in]{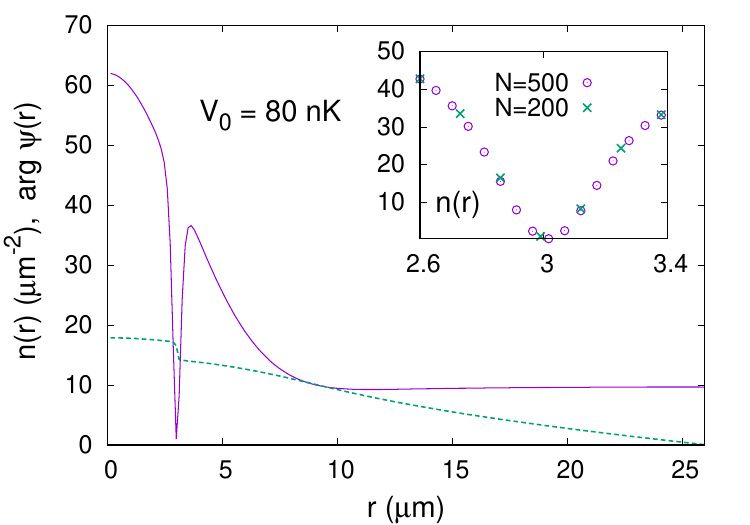}
\end{center}
\caption{\small
Profiles of the density (solid line) and the phase of $\psi$ (dashed line) for the new stationary 
solution. The inset shows detail of the density profile near the minimum, allowing for an estimate
of the minimum density (see text) and comparison of 
results from two different grid sizes.} 
\label{fig:soliton}
\end{figure}

The stationary solution itself is shown in Fig.~\ref{fig:soliton}. Note that the dip in the 
density is much larger than that for the droplet and occurs at a smaller radius. One may
observe similarity of the density profile with that of the ring dark soliton of the
conservative GP equation \cite{Theocharis&al}. We interpret our solution as a bound state 
of a ring soliton in the central region. The limit cycle into which
it decays if perturbed corresponds to small radial
oscillations of the soliton, accompanied by radiation of waves to infinity.

A comparison of results from two grid sizes in the inset of Fig.~\ref{fig:soliton} shows that
the dip, sharp as it is, is well resolved in our numerical computation.
The grid points for $N=500$ happen to fall almost symmetrically about the minimum of $n(r)$, 
which allows us to estimate density at the minimum as $n_{\min} = 0.37~\mic^{-2}$. 
This estimate is confirmed by computation on an even finer grid ($N=1000$).

Using the NR method, we have found stationary solutions of this kind for a wide range of potential
depths: from $V_0$ as low as 19.6 nK to as high as $V_0 = V_{c2} \approx 94.6$ nK. 
Note that this range includes
the critical depth $V_{c1}$ of the preceding section. 
The way the new solution disappears at $V_0 = V_{c2}$ is similar to how 
the ground state disappears at $V_0 = V_{c1}$: 
in both cases, a small supersonic region forms near $r = 10~\mic$.

In addition to having an $\ell = 0$ unstable mode for certain ranges of $V_0$,
the second stationary solution always (as far as we can tell) has several unstable modes with 
$\ell \geq 2$, with the most unstable one for $V_0$ close to $V_{c2}$ corresponding to
$\ell = 9$. We interpret these as evidence of a transverse (snaking) instability of the type
discussed previously \cite{Theocharis&al,Tamura&al}
for ring dark solitons in the conservative system without flow.

To become operational, a transverse instability requires preexisting fluctuations with $\ell \neq 0$,
such as those due to thermal or quantum noise. 
The smaller these initial fluctuations are, the longer it will take for the instability to develop. 
This means that results of 
numerical evolution of a rotationally symmetric problem can remain applicable for quite some
time, provided initial fluctuations with $\ell \neq 0$ are small enough. Indeed, for the system
without flow, extended intervals
preceding transverse fragmentation of ring solitons have been observed both 
in fully 2D simulations of the GP equation with a realistic amount of noise and in the experiment 
\cite{Tamura&al}.

\section{Resistive states}

Results of the preceding section may suggest that a quench of the potential to a depth above 
$V_{c1}$ but below $V_{c2}$ will lead directly to the new superfluid
state. As we will now see, that is indeed observed in numerical evolution for weak 
quenches.\footnote{
An oscillating soliton of the type described in Sec.~\ref{sec:sfsol} 
has been observed also in simulations of the time-dependent 
Eq.~(\ref{eq}) in which the potential is ramped gradually from zero, rather than quenched
(C.-L. Hung, private communication).}
A stronger quench, however,
leads instead to complex quasi-periodic dynamics of the type described in Ref.~\cite{self-osc}.

\begin{figure}
\begin{center}
\includegraphics[width=3.25in]{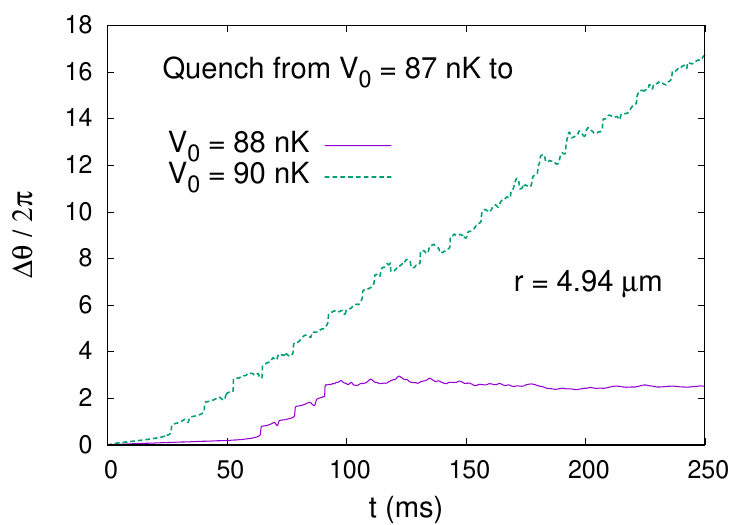}
\end{center}
\caption{\small
Change of the phase of the order parameter (in units of $2\pi$)
as a function of time at a given radius for
two instantaneous quenches of the potential to supercritical values, 
with the system initially in a superfluid ground state.
}
\label{fig:phase}
\end{figure}

To further characterize that dynamics, 
we compute the change over time of the phase $\theta = \arg \psi$ at some
radius $r$ in the ``active" region (where the flow speed reaches large values).
This is done by computing the time derivative of $\theta$  at radius $r$
from the equation
\be
\hbar \dot{\theta} = \frac{\hbar^2}{2m} \left( \mbox{Re} \frac{\nabla^2 \psi}{\psi} - 2 g n \right)
- V(r) + \mu 
\label{dott}
\ee
and integrating it over time to obtain
\be
\D \theta(t) = \int_0^t \dot{\theta}(t') dt' \, .
\label{theta}
\ee
Results shown in Fig.~\ref{fig:phase} are for $\D \theta(t)$ at $r = 4.94~\mic$ for
two instantaneous quenches, both from the ground state
at $V_0 = 87$ nK, but one to $V_0 = 88$ nK, which is just above $V_{c1}$, and the other
to a larger $V_0 = 90$ nK.
We see that for the weaker quench, after three well-defined phase slips at $t < 100$ ms, the phase
settles to a nearly constant value. Eventually, it approaches a limit cycle of the type discussed
in Sec.~\ref{sec:sfsol}. On the other hand, for the stronger quench, phase growth 
continues unabated, at a more or less constant rate but with a noticeable quasi-periodic component.

By analogy with Josephson's relation for superconductors, we can think of $\dot{\theta}$ or, more 
precisely, its average over a long period of time as a ``voltage," whose zero (nonzero) 
values correspond to superfluid (resistive) states. For the superfluid ground state, 
$\dot{\theta} = 0$. We extend
definition of superfluidity to states, like the limit cycle developing for $V_0 = 88$ nK in 
Fig.~\ref{fig:phase}, for which $\dot{\theta}$ is nonzero but
averages to zero over asymptotically long time.
States with a nonzero and approximately constant time average of $\dot{\theta}$, on the other hand, 
can be described in this terminology as resistive.

\section{Discussion}
Repeated phase slips, accompanying emission of solitons, occur also in the conservative
1D GP equation \cite{Hakim}. A peculiarity of the present (dissipative 2D) case 
is the complex quasi-periodic nature of the process at its later stages (cf. Fig.~\ref{fig:phase}). 
Let us note that this peculiarity
may in fact be related to another attribute of the present case that we have highlighted 
here---the second 
superfluid solution that exists even after the first one (the ground state) has disappeared.

Indeed, imagine that instead of the phase portrait sketched in
Fig.~\ref{fig:portr}, we had one where the system relaxed to the ground state for either direction of
the initial displacement. Then, at 
the critical $V_0 = V_{c1}$, at which the ground state and the droplet merge, there would be an orbit
that starts and ends at the same point (a homoclinic orbit in the terminology of dynamical 
systems theory). By a theorem of Shilnikov \cite{Shilnikov,Gonchenko&al}, 
for $V_0$ just above $V_{c1}$, there would then be a stable
limit cycle in the vicinity of that orbit.  
Existence of the second superfluid solution, however,
obviates this argument: the second solution represents a different limit cycle, which lies far from
the original ground state/droplet locus. As it turns out, the domain of attraction of this solution
is quite small, allowing more complex dynamics to readily emerge.

\begin{acknowledgments}
The author would like to thank Chen-Lung Hung for discussions.
This work was supported in part by the DOE QuantISED program through a theory consortium 
at Fermilab and by the W. M. Keck Foundation.
\end{acknowledgments}


\begin{thebibliography}{99}
\bibitem{Landau} L. Landau,
 ``Theory of the Superfluidity of Helium II," Phys. Rev. {\bf 60}, 356 (1941).
 \bibitem{Unruh} W. G. Unruh, 
  ``Experimental Black-Hole Evaporation?" Phys. Rev. Lett. {\bf 46}, 1351 (1981).
\bibitem{Hawking}
  S. W. Hawking, 
  ``Particle Creation by Black Holes,"
  Comm. Math. Phys. {\bf 43}, 199 (1975).
\bibitem{Steinhauer} J. Steinhauer, 
  ``Observation of quantum Hawking radiation and its entanglement in an analogue black hole," 
  Nature Physics {\bf 12}, 959 (2016).
\bibitem{de_Nova&al} J. R. M. de Nova, K. Golubkov, V. I. Kolobov, and J. Steinhauer,
  ``Observation of thermal Hawking radiation and its temperature in an analogue black hole,"
  Nature {\bf 569}, 688 (2019).
\bibitem{Kolobov&al} V. I. Kolobov, K. Golubkov, J. R. M. de Nova, and J. Steinhauer,
  ``Observation of stationary spontaneous Hawking radiation and the time evolution of an analogue 
  black hole,"
  Nature Physics {\bf 17}, 362 (2021).
\bibitem{Tettamanti&al} M. Tettamanti, S. L. Cacciatori, A. Parola, and I. Carusotto, 
  ``Numerical study of a recent black-hole lasing experiment,"
  Europhys. Lett. {\bf 114}, 60011 (2016).
\bibitem{Wang&al} Y.-H. Wang, T. Jacobson, M. Edwards, and C. W. Clark, 
  ``Mechanism of stimulated Hawking radiation in a laboratory Bose-Einstein condensate,"
  Phys. Rev. A {\bf 96}, 023616 (2017).
\bibitem{self-osc} H. Tamura, S. Khlebnikov, C.-A. Chen, 
  and C.-L. Hung, 
 ``Observation of self-oscillating supersonic flow across an acoustic horizon in two dimensions,"
 arXiv:2304.10667.
\bibitem{Hakim} V. Hakim,
  ``Nonlinear Schr\"{o}dinger flow past an obstacle in one dimension,"
  Phys. Rev. E {\bf 55}, 2835 (1997).
\bibitem{Pavloff} N. Pavloff, 
  ``Breakdown of superfluidity of an atom laser past an obstacle," 
  Phys. Rev. A {\bf 66}, 013610 (2002).
\bibitem{Engels&Atherton} P. Engels and C. Atherton,
  ``Stationary and Nonstationary Fluid Flow of a Bose-Einstein Condensate Through a Penetrable 
  Barrier," 
  Phys. Rev. Lett. {\bf 99}, 160405 (2007).
\bibitem{SBT} W. J. Skocpol, M. R. Beasley, and M. Tinkham, 
 ``Phase-Slip Centers and Nonequilibrium Processes in Superconducting Tin Microbridges," 
 J. Low Temp. Phys. {\bf 16}, 145 (1974).
\bibitem{Michel&Parentani} F. Michel and R. Parentani, 
  ``Nonlinear effects in time-dependent transonic flows: An analysis of analog black hole stability," 
  Phys. Rev. A {\bf 91}, 053603 (2015).
\bibitem{deNova&al:2016} J. R. M. de Nova, S. Finazzi, and I. Carusotto, 
  ``Time-dependent study of a black-hole laser in a flowing atomic condensate,"
  Phys. Rev. A {\bf 94}, 043616 (2016).
\bibitem{lapack} The LAPACK Team, ``LAPACK---Linear Algebra PACKage," netlib.org/lapack/.
\bibitem{Little} W. A. Little, 
  ``Decay of Persistent Currents in Small Superconductors,"  
  Phys. Rev. {\bf 156}, 396 (1967).
\bibitem{LA} J. S. Langer and V. Ambegaokar, 
  ``Intrinsic Resistive Transition in Narrow Superconducting Channels,"
  Phys. Rev. {\bf 164}, 498 (1967).
\bibitem{Theocharis&al} G. Theocharis, D. J. Frantzeskakis, P. G. Kevrekidis, B. A. Malomed, 
  and Y. S. Kivshar,
  ``Ring Dark Solitons and Vortex Necklaces in Bose-Einstein Condensates,"
  Phys. Rev. Lett. {\bf 90}, 120403 (2003).
\bibitem{Tamura&al} H. Tamura, C.-A. Chen, and C.-L. Hung,
  ``Observation of Self-Patterned Defect Formation in Atomic Superfluids---from Ring Dark Solitons 
  to Vortex Dipole Necklaces,"
  Phys. Rev. X {\bf 13}, 031029 (2023).
\bibitem{Shilnikov} L. P. Shilnikov,
  ``Some cases of generation of period motions from singular trajectories,"
  Mat. Sb. {\bf 103}, 443 (1963).
\bibitem{Gonchenko&al} S. Gonchenko, A. Kazakov, D. Turaev, and A. L. Shilnikov,
  ``Leonid Shilnikov and mathematical theory of dynamical chaos,"
  Chaos {\bf 32}, 010402 (2022).
\end{thebibliography}
\end{document}